\documentclass[conference]{IEEEtran}
\IEEEoverridecommandlockouts

\usepackage{cite}
\usepackage{amsmath,amssymb,amsfonts}
\usepackage{algorithmic}
\usepackage{graphicx}
\usepackage{textcomp}
\usepackage{xcolor}
\usepackage{booktabs}
\usepackage{multirow}
\def\BibTeX{{\rm B\kern-.05em{\sc i\kern-.025em b}\kern-.08em
    T\kern-.1667em\lower.7ex\hbox{E}\kern-.125emX}}
\begin{document}

\title{Exploiting Layer-Specific Vulnerabilities to Backdoor Attack in Federated Learning}

\author{
	\IEEEauthorblockN{
		Mohammad Hadi Foroughi\IEEEauthorrefmark{1}\IEEEauthorrefmark{4},
		Seyed Hamed Rastegar\IEEEauthorrefmark{2}\IEEEauthorrefmark{4},
        Mohammad Sabokrou\IEEEauthorrefmark{3},
		Ahmad Khonsari\IEEEauthorrefmark{1}\IEEEauthorrefmark{4}
	}
    \IEEEauthorblockA{
    	\IEEEauthorrefmark{1}School of Electrical and Computer Engineering, University of Tehran, Tehran, Iran.\\
        \IEEEauthorrefmark{2}School of Electrical and Computer Engineering, Shiraz University, Shiraz, Iran.\\
		\IEEEauthorrefmark{3}Okinawa Institute of Science and Technology, Okinawa, Japan\\
        \IEEEauthorrefmark{4}School of Computer Science, Institute for Research in Fundamental Sciences (IPM), Tehran, Iran.\\
		Emails: \{mhforoughi, a\_khonsari\}@ut.ac.ir, h.rastegar@shirazu.ac.ir, mohammad.sabokrou@oist.jp}
	}

\maketitle

\begingroup
\renewcommand\thefootnote{}
\footnotetext{
This paper has been accepted for publication in IEEE ICC 2026.
© 2026 IEEE. Personal use of this material is permitted. Permission from IEEE must be obtained for all other uses, in any current or future media, including reprinting/republishing this material for advertising or promotional purposes, creating new collective works, for resale or redistribution to servers or lists, or reuse of any copyrighted component of this work in other works. 
}
\endgroup

\begin{abstract}
Federated learning (FL) enables distributed model training across edge devices while preserving data locality. This decentralized approach has emerged as a promising solution for collaborative learning on sensitive user data, effectively addressing the longstanding privacy concerns inherent in centralized systems. However, the decentralized nature of FL exposes new security vulnerabilities, especially backdoor attacks that threaten model integrity. To  investigate this critical concern, this paper presents the Layer Smoothing Attack (LSA), a novel backdoor attack that exploits layer-specific vulnerabilities in neural networks. First, a Layer Substitution Analysis methodology systematically identifies backdoor-critical (BC) layers that contribute most significantly to backdoor success. Subsequently, LSA strategically manipulates these BC layers to inject persistent backdoors while remaining undetected by state-of-the-art defense mechanisms. Extensive experiments across diverse model architectures and datasets demonstrate that LSA achieves a remarkably backdoor success rate of up to 97\% while maintaining high model accuracy on the primary task, consistently bypassing modern FL defenses. These findings uncover fundamental vulnerabilities in current FL security frameworks, demonstrating that future defenses must incorporate layer-aware detection and mitigation strategies.
\end{abstract}


\section{Introduction}
The rapid increase in Internet of Things (IoT) devices has created a massive amount of data at the network's edge. Federated Learning (FL) has emerged as a transformative paradigm for leveraging this data, enabling collaborative model training across millions of distributed IoT devices—from smart home sensors to industrial control systems—without compromising user privacy \cite{fl}. By keeping raw data localized and only sharing model updates, FL is instrumental in privacy-sensitive IoT applications such as personalized healthcare, smart city management, and critical infrastructure monitoring \cite{fl-iot}.

Despite its privacy advantages, the decentralized nature of FL creates significant security challenges. The central server in an FL system has no direct visibility into the local training processes of participating devices, making it susceptible to malicious actors who can inject poisoned model updates to corrupt the global model. Among the most insidious threats are backdoor attacks, where an adversary embeds a hidden trigger into the model. The compromised model performs normally on standard inputs but misclassifies inputs containing the specific trigger to a target class chosen by the attacker \cite{badnets,dba}. Such attacks are particularly dangerous in IoT environments, where they could be used to manipulate autonomous systems, disable security protocols, or cause widespread service disruptions\cite{badnets,flare}.

Existing research has proposed various backdoor attack strategies and corresponding defense mechanisms \cite{how-backdoor-fl,flame}. However, many of these approaches treat the neural network as a monolithic black box, overlooking the fact that different layers may contribute unequally to the model's vulnerability. Recent work has suggested that the influence of a backdoor is often concentrated in a small subset of a model's layers \cite{lpa}. This observation opens a new attack surface where adversaries can craft more subtle and targeted attacks that are harder to detect.

This paper addresses this critical gap by exploring layer specific vulnerabilities in FL. In this work, we:
\begin{enumerate}
\item 	Introduce an enhanced Layer Substitution Analysis methodology to systematically identify backdoor-critical (BC) layers. This technique allows us to pinpoint the specific layers within a neural network that are most influential in the execution of a backdoor task. 
\item 	Propose a novel Layer Smoothing Attack (LSA), a sophisticated backdoor injection technique that exploits the identified BC layers. LSA strategically modifies only the BC layers and uses a smoothing technique to ensure compatibility with the rest of the model, making the malicious updates statistically similar to benign ones and thus evading detection. 
\item 	Provide a comprehensive empirical evaluation of LSA against state-of-the-art FL defense mechanisms across multiple model architectures and benchmark datasets. Our results demonstrate that LSA can successfully inject backdoors with high success rates while preserving model accuracy, highlighting a significant vulnerability in current FL security frameworks.
\end{enumerate}

Our findings underscore the urgent need for a new class of defense mechanisms that are aware of the model's internal structure and can counteract layer-specific threats. The remainder of this paper is organized as follows: Section II reviews related work. Section III details our methodology for identifying BC layers and the LSA technique. Section IV presents our experimental setup and results. Finally, Section V concludes the paper and discusses future research directions. 

\section{Related Work}
Backdoor attacks pose a significant threat to the integrity of FL systems. The seminal work on "BadNets" demonstrated the feasibility of injecting backdoors into neural networks by poisoning the training data \cite{badnets}. This concept was later adapted to the FL setting, where adversaries can poison their local datasets to corrupt the global model \cite{byzantine-tolerant}. Bagdasaryan et al. \cite{how-backdoor-fl} showed that a single malicious client could inject a backdoor by scaling their model update, a technique known as a model replacement attack. More advanced attacks have since been developed. The Distributed Backdoor Attack (DBA) distributes the backdoor trigger across multiple colluding attackers, making individual updates appear benign and thus harder to detect \cite{dba}. These attacks demonstrate a clear trend towards more stealthy and sophisticated methods of compromising FL systems.

In response, a variety of defense mechanisms have been proposed. These can be broadly categorized into two groups: robust aggregation rules and anomaly detection. Robust aggregation methods, such as Multi-Krum \cite{multikrum} and Trimmed Mean \cite{trimmed-mean}, aim to filter out malicious updates during the aggregation process at the central server. Anomaly detection techniques, such as FLAME \cite{flame}, analyze the statistical properties of model updates to identify and reject those that deviate significantly from the norm.

However, a key limitation of many existing defenses is their "black-box" approach, which analyzes model updates as a whole without considering their internal structure. Zhuang et al. \cite{lpa} recently showed that attacks targeting specific layers can be highly effective and can bypass defenses that focus on global model properties. Their Layer-wise Poisoning Attack (LPA) demonstrated that manipulating only a few critical layers is sufficient to inject a backdoor. Our work builds upon this insight by developing a more systematic method for identifying these critical layers and a more advanced attack technique that is specifically designed to evade modern sophisticated defenses.

\section{Methodology}
\subsection{Federated Learning}
Federated learning leverages a large set of distributed users, denoted as $U$, to train a global model iteratively without transferring private user data to a central server. The objective in FL is to solve an optimization problem, as described in \eqref{federated_optimazation}. Where for each user $i$, $f_ i(w^{(i)})$ is an local objective function, $D^{(i)} $ is local dataset, $l(x,y;w^{(i)})$ is loss function. The training process involves collecting local updates from participants and aggregating them to iteratively refine the global model.

\begin{equation}
\label{federated_optimazation}
\begin{gathered}
\min _w F(w):=\sum_{i \in \mathcal{N}} p^{(i)} f_i\left(w^{(i)}\right) \\
f_i\left(w^{(i)}\right)=\frac{1}{\left|D^{(i)}\right|} \sum_{(x, y) \in D^{(i)}} \ell\left(x, y ; w^{(i)}\right) \\
p^{(i)}=\frac{\left|D^{(i)}\right|}{\sum_{i \in \mathcal{N}}\left|D^{(i)}\right|}
\end{gathered}
\end{equation}

In backdoor attacks on federated learning, an adversary modifies the global model to misclassify specific inputs with a trigger while ensuring it performs well on clean data. This means the model behaves normally on standard inputs but exhibits malicious behavior when presented with poisoned samples.

We assume the adversary controls a subset of malicious users, denoted as $m = \{ 1,...,M \}$. In this scenario, we consider fewer than half of the users to be malicious. If most users were compromised, existing federated learning defenses would have difficulty countering the attack effectively. As suggested in previous studies \cite{flame,how-backdoor-fl,trimmed-mean,rlr,sybil,little,Byzantine-Robust} adversarial users can collaborate to refine their attack strategies through communication. Moreover, the adversary has continuous access to the global model during each training round, enabling them to adjust model weights and manipulate datasets used by malicious participants \cite{3dfed,Byzantine-Robust}.

\subsection{Identify Backdoor Critical layers}
\begin{figure*}[htbp]
\centerline{\includegraphics{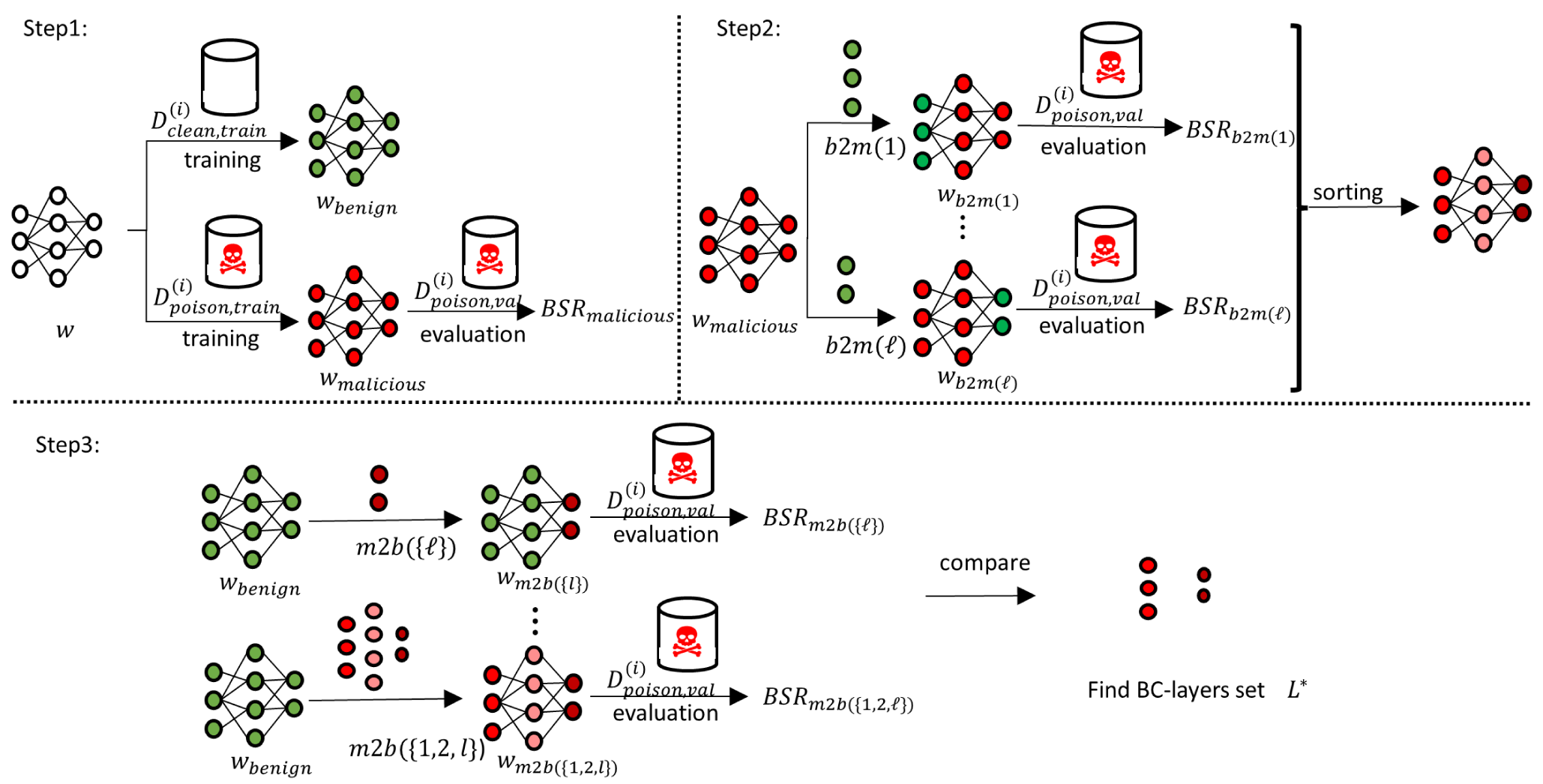}}
\caption{The layer substitution process for identifying BC-layers in LSA attack}
\label{layer-substitution}
\end{figure*}

In an FL setting, a global model $w$ is trained collaboratively by multiple participants. Each training round, participants receive the latest global model, update it using their local dataset, and send back refined parameters for aggregation. As a result, if some participants are malicious and trained on poisoned datasets, there is an opportunity to distinguish between models that are trained on clean data versus those are trained on poisoned data.
The core idea of this paper is that certain layers, known as BC-layers, play a crucial role in backdoor attacks. Replacing a sensitive layer in a benign model with the corresponding layer from a malicious model significantly reduces backdoor accuracy. Conversely, transferring these critical layers from a malicious model to a benign one increases its backdoor success rate (BSR) to a level similar to the malicious model. The following steps outline how to identify BC-layers, as illustrated in Fig. \ref{layer-substitution}.

Step 1:
At each FL round, a malicious user $i$ receives the global model $w$. The user partitions its local dataset $D^{(i)}$ into clean and poisoned subsets for training and validation: $D_{(clean,train)}^{(i)}$, $D_{(poison,train)}^{(i)}$, $D_{(clean,val)}^{(i)}$ and $D_{(poison,val)}^{(i)}$. 
The poisoned subsets include the backdoor trigger. First, the malicious user trains $w$ on $D_{clean,train}^{(i)}$ to convergence, producing a benign local model, $w_{benign}$.
Second, malicious user fine-tunes $w$ on $D_{poison,train}^{(i)}$ to create a malicious local model, $w_{malicious}$. The effectiveness of this model is evaluated on $D_{poison,val}^{(i)}$ to establish a baseline BSR, $BSR_{malicious}$.

Step 2:
To measure the importance of each layer $l$ to the backdoor, we substitute the $l$-th layer of the malicious model $w_{malicious}$ with the corresponding layer from the benign model $w_{benign}$. We then evaluate the BSR of this new model, $w_{b2m(l)}$. The drop in BSR, denoted as $\Delta BSR_{b2m(l)}$, indicates the contribution of the $l$-th layer to the backdoor as defined in Equation \eqref{diff_bsr}. Layers are then sorted in descending order based on their $\Delta BSR$ value.

\begin{equation}
\label{diff_bsr}
 \Delta BSR_{b2m(l)} =BSR_{malicious} - BSR_{b2m(l)} 
\end{equation}

Step 3:
Starting from the sorted list, iteratively build a set $L^*$ of candidate critical layers. Let $w_{m2b(L^*)}$ be the benign model with layers in $L^*$ replaced by their malicious counterparts. After each insertion, evaluate $BSR_{w_{m2b(L^*)}}$ on $D^{(i)}_{(poison,val)}$ and stop when $BSR_{m2b(L^*)} \ge \tau BSR_{malicious},\ \tau\in(0,1]$ which identifies a compact set of BC-layers. This captures both single-layer and multi-layer dependencies.

\subsection{Poisoning}
Identifying BC-layers provides an opportunity to develop more precise and stealthy backdoor attacks in federated learning. This work introduces the Layer Smoothing Attack (LSA), an approach that selectively poisons BC-layers with minimal perturbations. By applying subtle changes to specific layers, LSA ensures that the attack remains undetectable while still achieving a high backdoor success rate. This strategy allows attackers to circumvent state-of-the-art defense mechanisms, making it a particularly effective method in federated learning environments. The impact and effectiveness of LSA are further evaluated in Section V.

In the $t-th$ round of FL, a subset of malicious users conducts a layer substitution process to identify BC-layers $L^*$. Substituting layers from the malicious model into the benign model often degrades both main-task accuracy and BSR. A key reason is the data distribution mismatch between the two models. Malicious layers are trained on poisoned data, incorporating features or triggers absent in clean datasets. When these malicious layers are inserted into a model trained on benign data, they can no longer rely on the same input distributions and internal feature representations they were designed for. This disconnect produces inconsistencies throughout the network, disrupting both the clean-task performance and the backdoor functionality.

Furthermore, layer interdependence plays a crucial role, as each layer’s function depends on the outputs of preceding layers. Replacing even a single layer can disrupt this balance, leading to inconsistencies in the model’s learned parameters.

To resolve these issues, a fine-tuning phase is essential. By freezing non-critical layers and allowing only the newly inserted malicious layers to update, the model can gradually resolve parameter inconsistencies. This approach offers two key benefits: it retains the benign model’s previously learned knowledge while enabling the malicious layers to adapt their weights. To achieve this, in \(w_{m2b(L^*)}\) we freeze all layers in \(L-L^*\), and then we fine-tune the updated model on the poisoned training data \(D^i_{(\mathrm{poison},\mathrm{val})}\) for a predefined number of epochs. This allows the newly substituted malicious layers to adjust their weights in harmony with the frozen ones; we refer to the resulting model as \(w_{ft}\).

To effectively manipulate the global model, the attacker defines a selection vector $v=[v_1,v_2,\ldots,v_l ]$ that determines which layers to use. If $v_j=0$, the $j$-th layer of the benign model $w_{benign}$ is used and if $v_j=1$, the $j$-th layer of the fine-tuned malicious model $w_{ft}$ is used.
To formalize the attack, the adversary’s goal at round $t$ is to optimize the malicious model so that it successfully embeds a backdoor trigger into the global model while remaining undetected. The attack objective is formulated as equation \eqref{poisoning}.

Where $w_{(t+1)}$ represents the global model weights at round $t+1$, and $A$ is the server’s aggregation function. To make the attack stealthy, malicious updates are carefully crafted to avoid detection by robust defense mechanisms. A critical component of LSA is evasion via model weight approximation. To avoid detection, malicious users carefully adjust their model weights to resemble those of benign users. Inspired by prior works \cite{Byzantine-Robust,Oblivion}, the adversary estimates benign user models by leveraging local benign updates stored within compromised devices. Using this knowledge, the attacker modifies its poisoned model updates to be statistically similar to benign models. To regulate the stealthiness of the attack, we introduce a hyperparameter $\lambda$ that controls the attack’s strength. Specifically, when $\lambda \ge 1$, the attack behaves similarly to a scaling attack, where small perturbations gradually amplify their influence over multiple training rounds. Additionally, we use a ReLU-based activation function to constrain the modifications applied to BC-layers.

\begin{equation}
\label{poisoning}
\begin{gathered}
\max _v \frac{1}{\left|D^{(i)}\right|} \sum_{(x, y) \in D^{(i)}} P\left[G\left(x^{\prime}\right)=y^{\prime} ; w_{t+1}\right], \\
\text { s.t. } w_{t+1}=\mathcal{A}\left(\widetilde{w}^{(1)}, \ldots, \widetilde{w}^{(M)}, w^{(M+1)}, \ldots, w^{(N)}\right), \\
\widetilde{w}^{(i)}=\lambda v \circ u_{\text {ft }}^{(i)}+\operatorname{ReLU}(1-\lambda) . v \circ u_{\text {a }}+(1-v) \circ u_{\text {a }} \\
u_{\text {a }}=\frac{1}{M} \sum_{k=0}^M u_{\text {benign }}^{(k)}
\end{gathered}
\end{equation}

Unlike naive poisoning approaches that modify BC-layers in every training round, LSA employs an adaptive poisoning strategy to optimize efficiency. Instead of continuously modifying layers, the attack periodically updates its target layers—e.g., once every five rounds—reducing detectability and significantly saving time.

\section{Evaluation and results}
To assess the performance of FL and the impact of backdoor attacks, we designed and implemented a simulation environment. The goal of this simulation is to analyze how the system behaves in the presence of malicious users and to evaluate its stability and accuracy under different conditions.

This simulation was implemented using the PyTorch framework and executed on an NVIDIA RTX 4060 GPU. In this setup, we considered 100 users participating in federated learning, with 10\% of them being malicious. In each training round, 10\% of the total users were randomly selected for local training. These parameters were chosen to reflect realistic federated learning scenarios where a subset of users may attempt to compromise the global model.

\subsection{Datasets}
To evaluate the impact of backdoor attacks, we used two widely recognized datasets for federated learning and machine learning research: CIFAR-10, containing 60,000 color images, and Fashion-MNIST, with 70,000 grayscale images of clothing items.

\subsection{Data Distribution}
The non-independent and non-identically distributed (non-IID) datasets were created by dividing users into $X$ groups, each corresponding to one of the $X$ classes in the dataset. The probability that a sample with label $y$ is assigned to group $x$ is denoted by $q$, while the probability of it being assigned to any other group is given by $(1-q)/(X-1)$. The samples within each group are uniformly distributed among the users in that group. A higher value of $q$ indicates a greater degree of non-IID distribution. Following previous studies \cite{lpa, fltrust, Byzantine-Robust}, the default value of $q$ in our simulations is set to 0.5.

\subsection{Model Architectures}
To evaluate the effectiveness of our proposed method, we conducted experiments using three different neural network architectures: 5-layer CNN \cite{lpa,fltrust,fldetector},(used for Fashion-MNIST), ResNet-18 \cite{resnet} (used for CIFAR-10), VGG-19 \cite{vgg} (used for CIFAR-10).
The architecture of the 5-layer CNN is detailed in Table \ref{tab:cnn-arch}.

\subsection{Evaluation Metrics}
We assess robustness using four metrics: Accuracy (ACC), measuring performance on clean test data; Backdoor Success Rate (BSR), indicating the likelihood of poisoned samples being classified into the attacker’s target class; Benign Acceptance Rate (BAR), showing the percentage of clean updates accepted by the defense; and Malicious Acceptance Rate (MAR), representing how often poisoned updates bypass the defense.

\subsection{Results}
To evaluate the stealthiness of the LSA against defense mechanisms, we measured the MAR and BAR of various attack methods when tested against the Flame defense mechanism. The results, presented in Table \ref{tab:mar&bar}, demonstrate that LSA consistently achieves a high MAR across all model architectures, indicating its ability to effectively bypass the defense mechanism while remaining undetected.

\begin{table}[htbp]
\caption{5-layer CNN model architecture}
\begin{center}
  \begin{tabular}{l c}
    \hline
    \textbf{Layer} & \textbf{Size} \\
    \hline
    Input                    & $28\times 28 \times 1$ \\
    Convolution + ReLU       & $3\times 3 \times 32$ \\
    Convolution + ReLU       & $3\times 3 \times 64$ \\
    Max Pooling              & $2\times 2$ \\
    Dropout                  & $0.5$ \\
    Fully Connected + ReLU   & $128$ \\
    Dropout                  & $0.5$ \\
    Dropout                  & $0.5$ \\
    Fully Connected          & $10$ \\
    \hline
  \end{tabular}
\label{tab:cnn-arch}
\end{center}
\end{table}

To evaluate the effect of the LSA attack on the primary learning task, experiments were conducted under two different conditions: a standard FL process without any attack, and an FL process under an LSA attack. The results, presented in Table \ref{tab:attack&no}, compare the accuracy of the main task in the ResNet-18 model between these two scenarios. The findings indicate that LSA does not significantly degrade the accuracy of the main task. This suggests that LSA successfully injects the backdoor without noticeably affecting the model’s primary classification performance, making it highly effective at maintaining stealth while executing the attack.

To gauge the effectiveness of backdoor attacks under various defense strategies, we conducted experiments on ResNet-18, VGG-19, and a 5-layer CNN. The results are summarized in Table \ref{tab:best-run-all-models}.

Across all defenses, LSA consistently achieves high BSR while preserving a competitive accuracy on the main task. In particular, under the FLAME defense—recognized as a robust mechanism—LSA attains a BSR of 96.98\% alongside a 72.19\% accuracy on the primary classification objective. In comparison, traditional attacks such as BadNets and DBA achieve BSR values of only 7.22\% and 3.88\%, respectively, when FLAME is in effect. These results underscore LSA’s substantial improvement over conventional backdoor methods in the ResNet-18 setting.

A similar pattern emerges with VGG-19, which also proves highly vulnerable to LSA. Under the FedAVG defense, LSA not only achieves a 96\% BSR but also maintains a main-task accuracy of 76.56\%, showcasing its effectiveness in poisoning larger and more complex models. Even against more sophisticated defenses, LSA sustains BSR values above 88\% in most cases. Notably, in the Multi-Krum defense, LSA reaches a BSR of 88.08\% with 82.23\% accuracy, whereas the LPA attack’s performance drops to a 56.81\% BSR and 69.37\% accuracy. These numbers highlight the consistent ability of LSA to evade defensive mechanisms, even under strict conditions.

\begin{figure}[htbp]
\centerline{\includegraphics[scale=0.5]{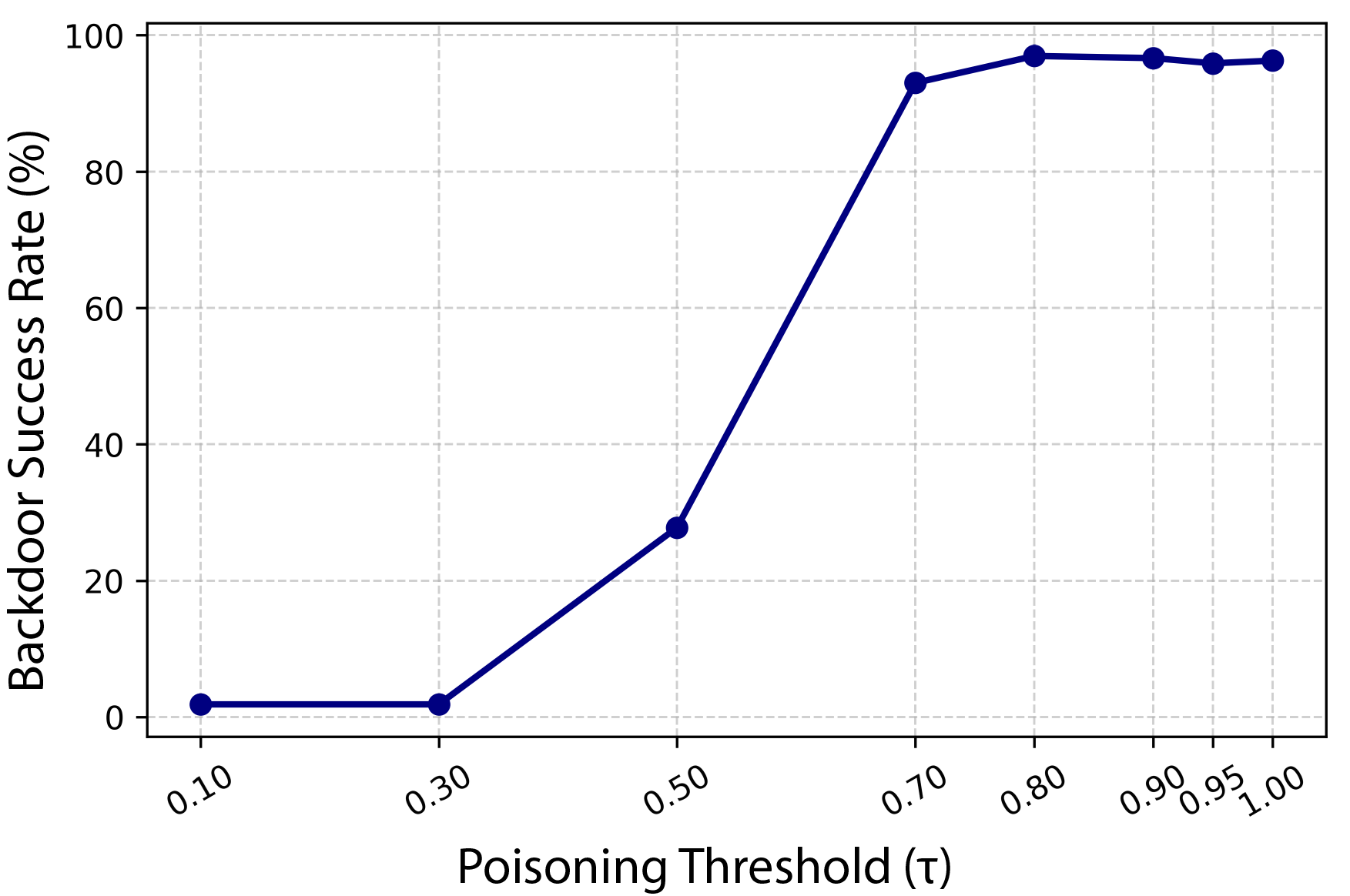}}
\caption{BSR vs. $\tau$ for ResNet18 under the Flame defense.}
\label{tau-plot}
\end{figure}

\begin{figure}[htbp]
\centerline{\includegraphics[scale=0.5]{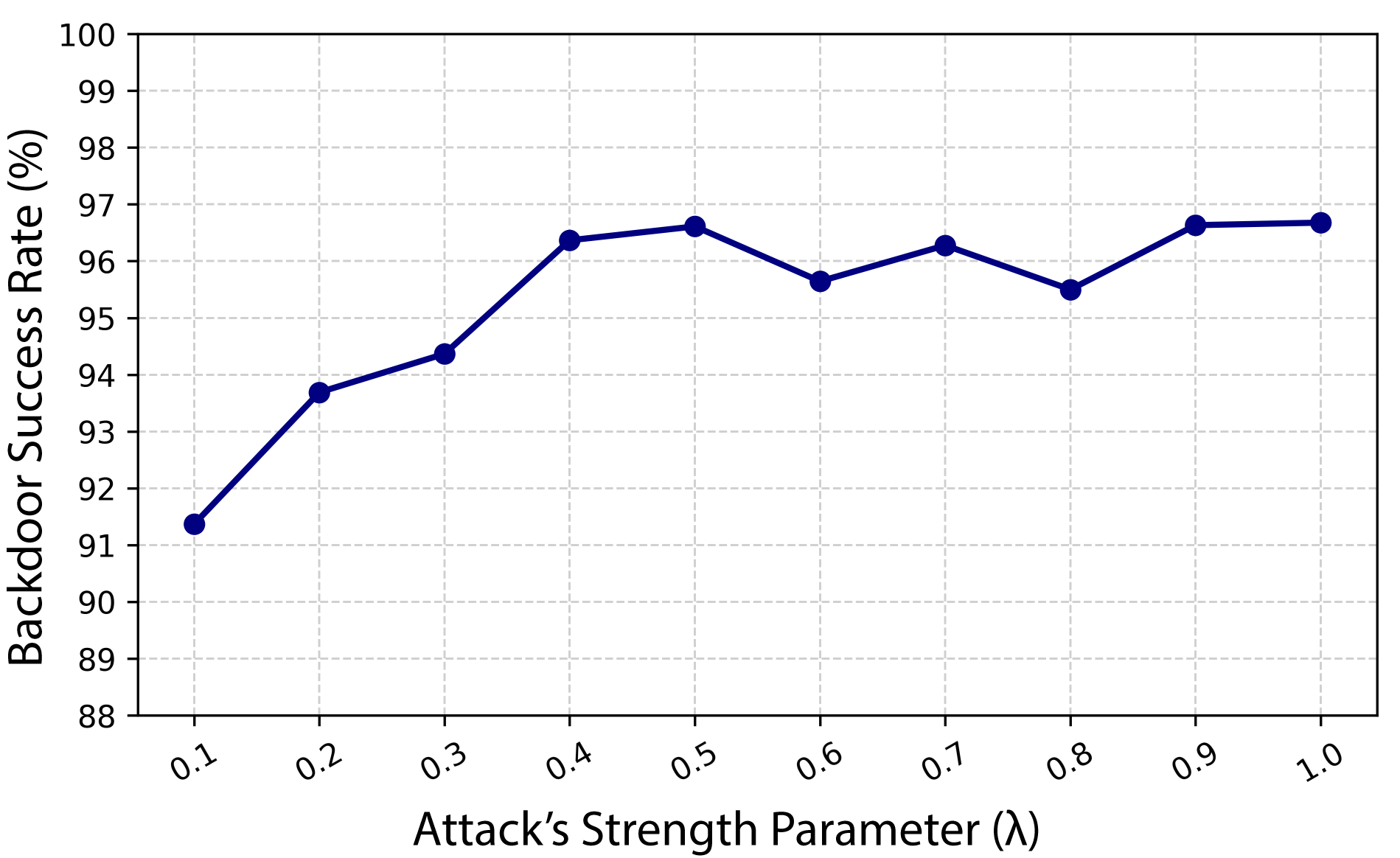}}
\caption{BSR vs. $\lambda$ for ResNet18 under the Flame defense.}
\label{lambda-plot}
\end{figure}

To further evaluate the impact of key hyperparameters, a series of experiments were conducted. Fig. \ref{tau-plot} and Fig. \ref{lambda-plot} illustrate the effect of two primary hyperparameters, $\tau$ and $\lambda$, on the BSR of the LSA attack when applied to the ResNet-18 model under the Flame defense mechanism.The results indicate that BSR increases with larger values of both parameters and reaches its maximum when $\lambda=1$ and $\tau=0.8$. Therefore, these values were used in all experiments as they provide the best balance between attack success and stealth.

\begin{figure}[htbp]
\centerline{\includegraphics[scale=0.5]{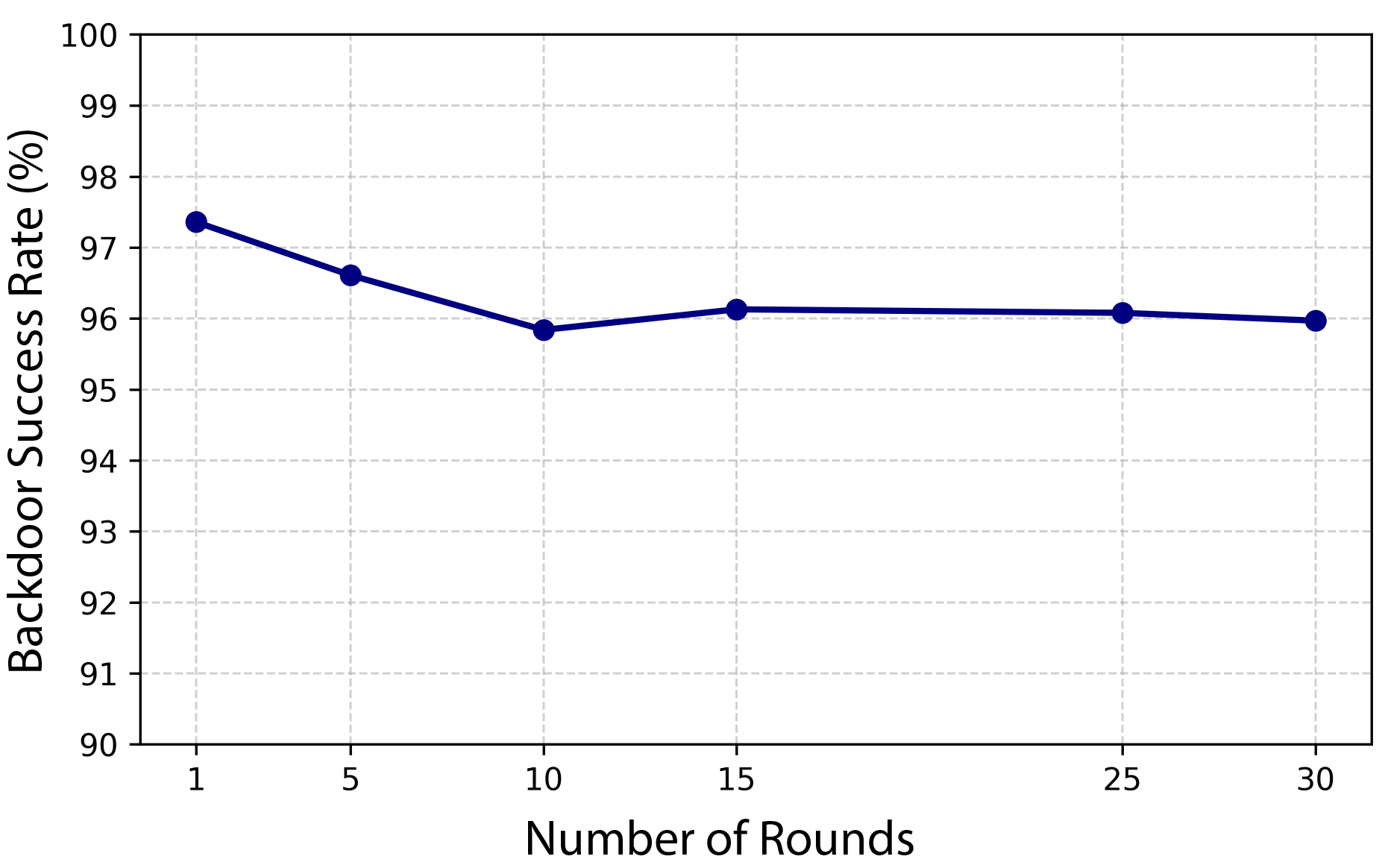}}
\caption{BSR vs. Detection period length for ResNet18 under the Flame defense.}
\label{period-lengths}
\end{figure}

We also examined the temporal frequency of identifying BC-layers to reduce the overall computational cost of the attack. Since the layer replacement process is time-consuming, the identification of BC-layers layers can be performed at specific intervals and the identified layers reused for subsequent attack rounds. Fig. \ref{period-lengths} illustrates the effect of varying detection-period lengths on the BSR of the LSA attack under the Flame defense using the ResNet-18 architecture. As observed, increasing the detection period does not significantly affect attack performance because BC-layers generally remain consistent throughout the FL process. Therefore, performing layer detection at fixed intervals yields substantial time savings without compromising attack effectiveness.

\begin{table}[htbp]
\caption{Accuracy of the Flame defense in detecting malicious and benign models (\%). Highest MAR and BAR values are bolded.}
\begin{center}
\begin{tabular}{l l c c}
    \hline
    \textbf{model} & \textbf{attack} & \textbf{MAR} & \textbf{BAR} \\
    \hline
    \multirow{4}{*}{ResNet18} & LSA     & \textbf{89.50} & 58.56 \\
                              & LPA     & 57.00 & 63.94 \\
                              & BadNets &  5.58 & 74.86 \\
                              & DBA     &  3.50 & \textbf{75.06} \\
    \hline
    \multirow{4}{*}{VGG19}    & LSA     & \textbf{97.00} & 55.89 \\
                              & LPA     & 65.50 & 59.77 \\
                              & BadNets & 16.58 & 73.54 \\
                              & DBA     & 12.25 & \textbf{74.10} \\
    \hline
    \multirow{4}{*}{CNN}      & LSA     & \textbf{97.00} & 56.00 \\
                              & LPA     & 46.50 & 61.50 \\
                              & BadNets &  0.25 & 66.81 \\
                              & DBA     &  0.50 & \textbf{67.11} \\
    \hline
  \end{tabular}
\label{tab:mar&bar}
\end{center}
\end{table}

\begin{table}[htbp]
\caption{Comparison of main task accuracy without and under LSA attack on ResNet18 (\%). Highest values are bolded.}
\begin{center}
  \begin{tabular}{lcc}
    \hline
    \textbf{defense} & \textbf{attack-free} & \textbf{LSA} \\
    \hline
    FedAVG       & \textbf{77.26} & 75.18 \\
    Median       & 77.37 & \textbf{79.61} \\
    Trimmed-Mean & 77.02 & \textbf{78.09} \\
    Multi-Krum   & 73.49 & \textbf{78.52} \\
    FLTrust      & 74.61 & \textbf{75.24} \\
    Flame        & 71.10 & \textbf{72.76} \\
    \hline
  \end{tabular}
\label{tab:attack&no}
\end{center}
\end{table}

\begin{table*}[htbp]
\caption{Best-run ACC and BSR for ResNet18, VGG19, and CNN (\%). Highest values per defense are bolded.}
\begin{center}
\begin{tabular}{l l cc cc cc cc}
    \toprule
    \textbf{Model} & \textbf{Defense} &
    \multicolumn{2}{c}{\textbf{DBA}} &
    \multicolumn{2}{c}{\textbf{BadNets}} &
    \multicolumn{2}{c}{\textbf{LPA}} &
    \multicolumn{2}{c}{\textbf{LSA}} \\
    & &
    \textbf{ACC} & \textbf{BSR} &
    \textbf{ACC} & \textbf{BSR} &
    \textbf{ACC} & \textbf{BSR} &
    \textbf{ACC} & \textbf{BSR} \\
    \midrule
    \multirow{6}{*}{ResNet18}
      & FedAVG       & \textbf{77.99} & 10.94 & 77.58 & 70.53 & 72.97 & 84.84 & 73.86 & \textbf{96.67} \\
      & Median       & 78.87 & 12.16 & 78.91 & 66.13 & 78.30 & 91.36 & \textbf{79.61} & \textbf{93.25} \\
      & Trimmed-Mean & \textbf{78.34} & 13.88 & 78.13 & 67.50 & 76.18 & 90.00 & 78.09 & \textbf{95.58} \\
      & Multi-Krum   & 73.02 &  5.61 & 74.49 &  3.95 & 68.45 & 84.16 & \textbf{76.89} & \textbf{92.01} \\
      & FLTrust      & \textbf{77.51} & 15.11 & 75.72 & 75.84 & 64.95 & 83.00 & 69.98 & \textbf{95.94} \\
      & Flame        & 75.27 &  3.88 & \textbf{76.04} &  7.22 & 68.30 & 82.14 & 72.19 & \textbf{96.98} \\
    \midrule
    \multirow{6}{*}{VGG19}
      & FedAVG       & \textbf{78.97} & 25.88 & 78.89 & 74.69 & 75.90 & 89.71 & 76.56 & \textbf{96.00} \\
      & Median       & \textbf{80.13} & 29.77 & 79.26 & 70.52 & 78.63 & 93.15 & 77.91 & \textbf{94.91} \\
      & Trimmed-Mean & \textbf{80.44} & 28.92 & 79.11 & 69.41 & 77.66 & 89.63 & 76.63 & \textbf{94.01} \\
      & Multi-Krum   & 64.81 &  8.44 & 58.93 &  7.84 & 69.37 & 56.81 & \textbf{82.23} & \textbf{88.08} \\
      & FLTrust      & \textbf{75.11} & 15.88 & 75.10 & 67.30 & 74.01 & 68.05 & 73.34 & \textbf{71.45} \\
      & Flame        & 63.30 &  7.33 & 62.91 &  7.78 & \textbf{66.51} & 92.52 & 64.83 & \textbf{93.00} \\
    \midrule
    \multirow{4}{*}{CNN}
      & FedAVG       & 87.95 & 99.90 & 88.28 & 99.90 & \textbf{89.29} & 83.45 & 86.73 & \textbf{99.96} \\
      & Multi-Krum   & \textbf{97.58} &  0.10 & 87.31 &  0.39 & 90.48 & 70.84 & 90.58 & \textbf{75.32} \\
      & FLTrust      & 89.31 & \textbf{100.00} & 89.51 & 68.97 & \textbf{90.16} & 84.50 & 87.81 & 96.90 \\
      & Flame        & 87.89 &  0.40 & 87.78 &  0.10 & 90.37 & 65.57 & \textbf{90.44} & \textbf{66.28} \\
    \bottomrule
  \end{tabular}
\label{tab:best-run-all-models}
\end{center}
\end{table*}


\section{Conclusion}
This study examined security vulnerabilities in federated learning at the layer level and introduced a novel attack method that exploits these weaknesses. Our work focused on identifying backdoor-critical layers and leveraging them for an effective and stealthy attack. By systematically analyzing the impact of layer substitution between benign and malicious models, we implemented the Layer Smoothing Attack, a method designed to evade existing defense mechanisms while maintaining effectiveness. Extensive experiments on different neural network architectures demonstrated that the proposed attack method effectively bypasses existing defenses. The attack achieved high backdoor success rates while preserving the model’s accuracy on its primary task. It exhibited superior evasion capabilities compared to prior approaches, highlighting the significant security risks posed by layer-level vulnerabilities in federated learning systems. This study exposes a key vulnerability in federated learning—small manipulations of specific layers can compromise the global model. The findings highlight the need for stronger defenses against layer-targeted attacks and contribute to advancing the security and resilience of federated learning systems..

\vspace{12pt}

\end{document}